# Atomic resolution imaging of the two-component Dirac-Landau levels in a gapped graphene monolayer

Wen-Xiao Wang[1,2,§], Long-Jing Yin[1,2,§], Jia-Bin Qiao[1,2,§], Tuocheng Cai[3], Si-Yu Li[1,2], Rui-Fen Dou[1], Jia-Cai Nie[1], Xiaosong Wu[3], and Lin He[1,2,*]

[1] Department of Physics, Beijing Normal University, Beijing, 100875, People's Republic of China

[2] The Center of Advanced Quantum Studies, Beijing Normal University, Beijing, 100875, People's Republic of China

[3] State Key Laboratory for Artificial Microsctructure and Mesoscopic Physics, Peking University, Beijing 100871, China and Collaborative Innovation Center of Quantum Matter, Beijing 100871, China

[§]These authors contributed equally to this work.
* Email: helin@bnu.edu.cn

**The wavefunction of massless Dirac fermions is a two-component spinor. In graphene, a one-atom-thick film showing two-dimensional Dirac-like electronic excitations, the two-component representation reflects the amplitude of the electron wavefunction on the *A* and *B* sublattices. This unique property provides unprecedented opportunities to image the two components of massless Dirac fermions spatially. Here we report atomic resolution imaging of the two-component Dirac-Landau levels in a gapped graphene monolayer by scanning tunnelling microscopy and spectroscopy. A gap of about 20 meV, driven by inversion symmetry breaking by the substrate potential, is observed in the graphene on both SiC and graphite substrates. Such a gap splits the $n = 0$ Landau level (LL) into two levels, $0_+$ and $0_-$. We demonstrate that the amplitude of the wavefunction of the $0_-$ LL is mainly at the *A* sites and that of the $0_+$ LL is mainly at the *B* sites of graphene, characterizing the internal structure of the spinor of the $n = 0$ LL. This provides direct evidence of the two-component nature of massless Dirac fermions.**

Because of the bipartite honeycomb lattice[1-3], which has two distinct sublattices (denoted by *A* and *B*), the wavefunctions describing the low-energy excitations near the Dirac points, commonly called *K* and *K'*, in graphene monolayer are two-component spinors. The Dirac spinors of the two cones in graphene have the form as:

$$|K\rangle = \begin{pmatrix} \psi_{KA} \\ \psi_{KB} \end{pmatrix} = \frac{1}{\sqrt{2}} \begin{pmatrix} 1 \\ \pm i e^{-i\theta_\tau} \end{pmatrix}, \quad |K'\rangle = \begin{pmatrix} \psi_{K'B} \\ \psi_{K'A} \end{pmatrix} = \frac{1}{\sqrt{2}} \begin{pmatrix} 1 \\ \pm i e^{i\theta_\tau} \end{pmatrix}. \quad (1)$$

Here $\theta_\tau = \arctan\left(\frac{q_{\tau,y}}{q_{\tau,x}}\right)$ is defined as the angle of wave vector $\boldsymbol{q}_\tau \equiv (q_{\tau,x}, q_{\tau,y})$ in momentum space. The two-component representation, which resembles that of a spin in a mathematically way[1-4], corresponds to the projection of the electron wavefunction on the *A* and *B* sublattices.

A site energy difference 2Δ between the sublattices, generated by substrates, could break the inversion symmetry of graphene and lift energy degeneracy of the *A* and *B* sublattices[5,6], as shown in Fig. 1a. This effect generates a gap $\Delta E = 2\Delta$ at the Dirac points, which was previously observed for graphene monolayer on top of SiC[7], graphite[8-11], and hexagonal boron nitride[12,13]. The gap, usually ranging from 10 meV to several tens meV, can result in a valley contrasting Hall transport in graphene[5,13]. In the quantum Hall regime, the broken symmetry of the graphene sublattices shifts the energies of the *n* = 0 LL in the *K* and *K'* valleys in opposite directions and, therefore, splits the *n* = 0 LL into the 0. and 0₊ LLs (here $\lambda$ = +, - denote the *K* and *K'* valleys respectively)[8-11], as schematically shown in Fig. 1c. Generally, the wavefunctions of the LLs in graphene are given by[2,14,15]

$$|K_n\rangle = \begin{pmatrix} \psi_{KA}^n \\ \psi_{KB}^n \end{pmatrix} = \begin{pmatrix} \sin(\alpha_n/2)\phi_{|n|-1} \\ \cos(\alpha_n/2)\phi_{|n|} \end{pmatrix}, \quad |K_n'\rangle = \begin{pmatrix} \psi_{K'B}^n \\ \psi_{K'A}^n \end{pmatrix} = \begin{pmatrix} \cos(\alpha_n'/2)\phi_{|n|-1} \\ \sin(\alpha_n'/2)\phi_{|n|} \end{pmatrix}, \qquad (2)$$

where $\sin\alpha_n = \dfrac{\hbar\omega_B\sqrt{|n|}\,\mathrm{sgn}_-(n)}{\sqrt{(\hbar\omega_B\sqrt{|n|})^2 + \Delta^2}}$, $\cos\alpha_n = -\dfrac{\Delta\,\mathrm{sgn}_-(n)}{\sqrt{(\hbar\omega_B\sqrt{|n|})^2 + \Delta^2}}$,

$\sin\alpha_n' = \dfrac{\hbar\omega_B\sqrt{|n|}\,\mathrm{sgn}_+(n)}{\sqrt{(\hbar\omega_B\sqrt{|n|})^2 + \Delta^2}}$, $\cos\alpha_n' = -\dfrac{\Delta\,\mathrm{sgn}_+(n)}{\sqrt{(\hbar\omega_B\sqrt{|n|})^2 + \Delta^2}}$, $\mathrm{sgn}_\pm(n) = \begin{cases} +1 & (n>0) \\ \pm 1 & (n=0) \\ -1 & (n<0) \end{cases}$

and $\omega_B = \sqrt{2ev_F^2 B/\hbar}$ with $v_F$ the Fermi velocity and $B$ the magnetic field (here $\phi_n$ is the usual Landau-level wavefunction). For $n = 0$, only the second components of the spinors are nonzero and we have $|K_0\rangle = \begin{pmatrix} 0 \\ \phi_0 \end{pmatrix}$ and $|K_0'\rangle = \begin{pmatrix} 0 \\ \phi_0 \end{pmatrix}$. It indicates that we can detect the $0_+$ LL, *i.e.*, the spinor $|K_0\rangle$, only at $B$ sites and detect the $0_-$ LL, *i.e.*, the spinor $|K_0'\rangle$, only at $A$ sites. With the help of high energy and spatial resolution of scanning tunneling microscopy (STM) and spectroscopy (STS), it is therefore possible to image the two components of the spinors in atomic resolution.

Figure 2a shows a representative STM image of a graphene multilayer grown on SiC substrate[16-19]. An intensity imbalance between the $A$ and $B$ sublattices, as shown in the atomic image of Fig. 2b, indicates the inversion symmetry breaking by the substrate potential[7-11]. Our STS, as shown in Fig. 2c, and Raman measurements (see Supplementary Fig. S1) show decoupling behavior of the topmost graphene monolayer and the supporting substrate, which is in agreement with earlier transport[17,20] and spectroscopy measurements[7,21,22]. The spectrum recorded in the magnetic field of 8 T exhibits Landau quantization of massless Dirac fermions, as expected to be observed in a graphene monolayer[7-11,23], and the Fermi velocity is estimated to be $v_F = (0.73 \pm 0.03) \times 10^6$ m/s. A notable feature of the tunneling spectra recorded at 8 T is the splitting of the $n = 0$ peak and its sensitive dependency on the recorded positions, as shown in Fig. 2c. The split of the $n = 0$ peak, ~ 20 mV, is attributed to a gap caused by the inversion symmetry breaking by the substrate potential[8-11]. The tunneling spectrum gives direct access to the local density of states (DOS) of the surface. Therefore, the spectra recorded at different positions, as shown in Fig. 2c, indicate that the $0_+$ LL at about 45 mV is significant only at *B* sites and the $0_-$ LL at about 65 mV is pronounced only at *A* sites. Such a feature, which has not been explored experimentally before, reminds us characteristics of the internal structure of the two-component spinors of the $0_-$ and $0_+$ LLs [2,14]. We will demonstrate below that the splitting of the $n = 0$ LL is a direct consequence of its two-component nature.

Figure 3a and 3b show conductance maps at 8 T at the bias voltages of the $0_-$ and $0_+$ LLs, respectively. The maps reflect spatial distribution of the local DOS at the bias voltages. Both the maps exhibit triangular lattice, indicating pronounced asymmetry of the $0_+$ and $0_-$ LLs on the sublattices. However, there is a very important difference between the two maps. The bright spots in the conductance map of the $0_+$ LL correspond to the dark spots of the triangular lattice, *i.e.*, the *B* sites, in the STM image, whereas, the bright spots in the map of the $0_-$ LL correspond to the bright spots of the triangular lattice, *i.e.*, the *A* sites, in the STM image (Similar conductance maps are also observed in a gapped graphene sheet on graphite surface, see Supplementary Fig. S2). At a fixed energy, the local DOS at position *r* is determined by the wavefunctions according to $\rho(r) \propto |\psi(r)|^2$. Therefore, the maps in Fig. 3a and 3b reflect atomic resolution images of the two-component Dirac-Landau levels. Theoretically, the spinor of the $0_+$ ($0_-$) LL only has non-zero component at *B* (*A*) sites, which is qualitatively consistent with the observed large asymmetry of the $0_+$ and $0_-$ LLs on the sublattices.

At *n* = 0 LL, the broken inversion symmetry lifts the degeneracies of both the sublattices and the valleys, as shown in Fig. 1c. At *n* ≠ 0, the *K* and *K′* valleys are doubly degenerate in energy, whereas there is still a difference between the amplitudes of the two components in the spinors of the LLs, as described by Eq. (2). For *n* > 0 (*n* < 0), the amplitude of the *A*-site (*B*-site) component of the spinors in both the *K* and *K′* valleys is predicted to be slightly larger than that of the *B*-site (*A*-site) component. Such a feature has also been demonstrated explicitly in our experiment. Figure 3c and Fig. 3d show conductance maps at 8 T at the bias voltages of the *n* = +1 and *n* = -1 LLs, respectively. Both the maps exhibit triangular contrasting and, obviously, the amplitude of the *n* = +1 (*n* = -1) LL on the *A* (*B*) sites is stronger than that on the *B* (*A*) sites (Similar conductance maps are also observed in the gapped graphene sheet on graphite surface, see Supplementary Fig. S3). However, the asymmetry between the amplitudes of the *A*-site component and *B*-site component of the spinors for the *n* = +1 and *n* = -1 LLs is much weaker than that for the $0_+$ and $0_-$ LLs, as demonstrated in Fig. 3. Such an asymmetry will further decrease with increasing the value of |*n*| according to Eq. (2). Therefore, we obtain almost honeycomb contrasting in the conductance map recorded at the bias voltage of 170

mV (the value of $n$ at 170 mV is estimated to be 3), as shown in Fig. 4a. Theoretically, the value of $\sin^2(a_n/2)/\cos^2(a_n/2)$, which reflects the asymmetry between the amplitudes of the $A$-site component and $B$-site component of the spinors, is estimated to be about 1.12 for $n = 3$.

To further compare our experimental results with the theory, we plot vertical line-cuts of the conductance maps of the $0_+$, $0_-$, -1, +1, and +3 LLs along $A$ and $B$ atoms in Fig. 4b. The theoretical result showing amplitudes of these LLs on the sublattices is also shown in Fig. 4c, which reproduces the overall features of our experimental results quite well. However, there is still an obvious difference if we compare the experimental result with the theoretical one quantitatively. For example, the $A$-site component of the $0_+$ LL is predicted to be zero, however, we always observe a nonzero value at the $A$-site in the conductance map of the $0_+$ LL. Such a discrepancy is mainly attributed to the overlap of the $0_+$ and $0_-$ LLs. There is a finite linewidth of the $0_+$ and $0_-$ LLs and they are separated by only 20 mV, as shown in Fig. 2c. Therefore, when we carry out conductance mapping at the bias voltage of the $0_+$ LL, there is a slightly contribution from the $0_-$ LL, which results in the nonzero value at the $A$-site. A concomitant result of this effect is that there is always a weak peak of the $0_-$ LL (or the $0_+$ LL) when we measure the tunneling spectra at the $B$-site (or $A$-site), as shown in Fig. 2c.

The validity of our experimental result is further confirmed by performing similar measurements on a graphene monolayer without breaking the inversion symmetry, *i.e.*, $\Delta = 0$, on graphite surface. For this zero-gap graphene, we always obtain honeycomb contrasting in the conductance maps recorded at 8 T at different bias voltages, which is consistent with the prediction of Eq. (2) by taking the limit $\Delta \rightarrow 0$ (see Supplementary Fig. S4 for experimental data). Therefore, our experimental result demonstrates that the splitting of the $n = 0$ LL is a direct consequence of its two-component nature and we realize atomic resolution imaging of the two-component Dirac-Landau levels in gapped graphene sheets on different substrates. The ability to tune the amplitude of the two-component of the Dirac spinors may pave the way for manipulating spins in other Dirac systems such as topological insulators[4,24,25].

**REFERENCES:**

1. Castro Neto, A. H., Guinea, F., Peres, N. M. R., Novoselov, K. S., Geim, A. K., The electronic properties of graphene. *Rev. Mod. Phys.* **81**, 109-162 (2009).

2. Goerbig, M. O., Electronic properties of graphene in a strong magnetic field. *Rev. Mod. Phys.* **83**, 1193-1243 (2011).

3. Katsnelson, M. I., Novoselov, K. S., Geim, A. K. Chiral tunneling and the Klein paradox in graphene. *Nature Phys.* **2**, 620-625 (2006).

4. Fu, Y.-S., Kawamura, M., Igarashi, K., Takagi, H., Hannaguri, T., Sasagawa, T., Imaging the two-component nature of Dirac-Landau levels in the topological surface state of $Bi_2Se_3$, *Nature Phys.* **10**, 815-819 (2014).

5. Xiao, D., Yao, W., Niu, Q., Valley-contrasting physics in graphene: magnetic moment and topological transport. *Phys. Rev. Lett.* **99**, 236809 (2007).

6. Min, H., Hill, J. E., Sinitsyn, N. A., Sahu, B. R., Kleinman, L., MacDonald, A. H., Intrinsic and Rashba spin-orbit interactions in graphene sheets. *Phys. Rev. B* **74**, 165310 (2006).

7. Zhou, S. Y., Gweon, G.-H., Fedorov, A. V., First, P. N., de Heer, W. A., Lee, D.-H., Guinea, F., Castro Neto, A. H., Lanzara, A., Substrate-induced bandgap opening in epitaxial graphene. *Nature Mater.* **6**, 770-775 (2007).

8. Li, G., Luican, A., Andrei, E. Y., Scanning tunneling spectroscopy of graphene on graphite. *Phys. Rev. Lett.* **102**, 176804 (2009).

9. Du, X., Skachko, I., Duerr, F., Luican, A., Andrei, E. Y., Fractional quantum Hall effect and insulating phase of Dirac electrons in graphene. *Nature* **462**, 192 (2009).

10. Miller, D. L., Kubista, K. D., Rutter, G. M., Ruan, M., de Heer, W. A., First, P. N., Stroscio, J. A. Observing the quantization of zero mass carriers in graphene. *Science* **324**, 924-927 (2009).

11. Li, G., Luican-Mayer, A., Abanin, D., Levitov, L., Andrei, E. Y., Evolution of Landau levels into edge states in graphene. *Nature Commun.* **4**, 1744 (2013).

12. Hunt, B., Sanchez-Yamagishi, J. D., Young, A. F., Yankowitz, M., LeRoy, B. J., Watanabe, Taniguchi, T., Moon, P., Koshino, M., Jarillo-Herrero, P., Ashoori, R. C., Massive Dirac fermions and Hofstadter butterfly in a van der Waals Heterostructure. *Science* **340**, 1427-1430 (2013).

13. Gorbachev, R. V., Song, J. C. W., Yu, G. L., Kretinin, A. V., Withers, F., Cao, Y., Mishchenko, A., Grigorieva, I. V., Novoselov, K. S., Levitov, L. S., Geim, A. K.,

Detecting topological currents in graphene superlattices. *Science* **346**, 448-451 (2014).

14. Koshino, M., Ando, T., Anomalous orbital magnetism in Dirac-electron systems: role of pseudospin paramagnetism. *Phys. Rev. B* **81**, 195431 (2010).

15. Lado, J. L., Gonzalez, J. W., Fernandez-Rossier, J., Quantum Hall effect in gapped graphene heterojunctions. *Phys. Rev. B* **88**, 035448 (2013).

16. Berger, C., Song, Z., Li, T., Li, X., Ogbazghi, A. Y., Feng, R., Dai, Z., Marchenkov, A. N., Conrad, E. H., First, P. N., de Heer, W. A., Ultrathin epitaxial graphite: 2D electron gas properties and a route toward graphene-based nanoelectronics. *J. Phys. Chem. B* **108**, 19912-19916 (2004).

17. Wu, X., Li, X., Song, Z., Berger, C., de Heer, W. A., Weak antilocalization in epitaxial graphene: evidence for chiral electrons. *Phys. Rev. Lett.* **98**, 136801 (2007).

18. Cai, T., Jia, Z., Yan, B., Yu, D., Wu, X., Hydrogen assisted growth of high quality epitaxial graphene on the C-face of 4H-SiC. *Appl. Phys. Lett.* **106**, 013106 (2015).

19. Zhang, R., Dong, Y., Kong, W., Han, W., Tan, P., Liao, Z., Wu, X., Yu, D., Growth of large domain epitaxial graphene on the C-face of SiC. *J. Appl. Phys.* **112**, 104307 (2012).

20. Berger, C., Song, Z., Li, X., Wu, X., Brown, N., Naud, C., Mayou, D., Li, T., Hass, J., Marchenkov, A. N., Conrad, E. H., First, P. N., de Heer, W. A., Electronic confinement and coherence in patterned epitaxial graphene. *Science* **312**, 1191-1196 (2006).

21. Hicks, J., Sprinkle, M., Shepperd, K., Wang, F., Tejeda, A., Taleb-Ibrahimi, A., Bertran, F., Le Fevre, P., de Heer, W. A., Berger, C., Conrad, E. H., Symmetry breaking in commensurate graphene rotational stacking: comparison of theory and experiment. *Phys. Rev. B* **83**, 205403 (2011).

22. Hass, J., Varchon, F., Millan-Otoya, J. E., Sprinkle, M., Sharma, N., de Heer, W. A., Berger, C., First, P. N., Magaud, L., Conrad, E. H., Why multilayer graphene on 4H-SiC(0001) behaves like a single sheet of graphene. *Phys. Rev. Lett.* **100**, 125504 (2008).

23. Yin, L.-J., Li, S.-Y., Qiao, J.-B., Nie, J.-C., He, L., Landau quantization in graphene monolayer, Bernal bilayer, and Bernal trilayer on graphite surface. *Phys. Rev. B* **91**, 115405 (2015).

24. Hasan, M. Z., Kane, C. L., Colloquium: topological insulators. *Rev. Mod. Phys.* **82**, 3045-3067 (2010).

25. Qi, X.-L., Zhang, S.-C., Topological insulators and superconductors. *Rev. Mod. Phys.* **83**, 1057-1110 (2011).

**Acknowledgements**

This work was supported by the National Basic Research Program of China (Grants Nos. 2014CB920903, 2013CBA01603, 2013CB921701), the National Natural Science Foundation of China (Grant Nos. 11422430, 11374035, 11474022, 51172029, 91121012), the program for New Century Excellent Talents in University of the Ministry of Education of China (Grant No. NCET-13-0054), Beijing Higher

Education Young Elite Teacher Project (Grant No. YETP0238).

**Author contributions**

L.H. conceived and provided advice on the experiment, analysis, and theoretical calculation. W.X.W. and L.J.Y. performed the STM experiments. J.B.Q., S.Y.L., and W.X.W. analyzed the data and performed the theoretical calculations. T.C. and X.W. synthesized the graphene on SiC substrate. L.H. wrote the paper. All authors participated in the data discussion.

**Competing financial interests:** The authors declare no competing financial interests.

**Figure Legends**

**Figure 1 | Electronic band structure and Landau quantization in a gapped graphene monolayer. a,** Schematic diagram of a graphene monolayer with a staggered sublattice potential breaking the inversion symmetry. The *A*-site and *B*-site are denoted by blue and red balls, respectively. **b,** Energy spectrum of a graphene monolayer with broken inversion symmetry. **c,** Schematic Landau levels and DOS of a gapped graphene monolayer in the quantum Hall regime. Peaks in the DOS correspond to the Landau Levels $n_\lambda$ ($\lambda = +, -$ denote the *K* and *K'* valleys).

**Figure 2 | STM images and STS spectra of a gapped graphene monolayer. a,** A STM topography of graphene on SiC (000-1) terrace ($V_{\text{sample}}$= -600 meV and $I = 300$ pA). **b,** Zoom-in atomic-resolution topography obtained in the black frame in panel **a**.

The honeycomb structure is overlaid onto the STM image. The STM image shows a triangular contrast indicating the broken sublattice symmetry. Here we define the bright spots as the *A*-site atoms and the dark spots as the *B*-site atoms. **c,** d*I*/d*V* spectra obtained at different positions, as marked in panel **b**. The black arrow denotes the position of the Dirac point of the topmost graphene sheet in zero magnetic field. In the magnetic field of 8 T, the spectra exhibit Landau quantization of massless Dirac fermions, as expected in graphene monolayer. LL indices are marked. The $n = 0$ LL splits into two peaks and the intensity of the two peaks depends sensitively on the recorded positions.

**Figure 3 | Conductance maps of the gapped graphene monolayer at different energies. a,** The conductance map recorded at the bias voltage of the $0_-$ LL ($V_{sample}$ = 65.5 mV). **b,** The conductance map recorded at the bias voltage of the $0_+$ LL ($V_{sample}$ = 45 meV). **c,** The conductance map recorded at the bias voltage of the +1 LL ($V_{sample}$ = 108 meV). **d,** The conductance map recorded at the bias voltage of the -1 LL ($V_{sample}$ = -22 meV). The honeycomb structure of graphene and the atomic resolution STM image are overlaid onto the maps. The amplitude of the $0_-$ and +1 LLs on the *A*-site is much stronger than that on the *B*-site, whereas, the amplitude of the $0_+$ and -1 LLs on the *B*-site is much stronger than that on the *A*-site.

**Figure 4 | Internal structures of different LLs. a,** The conductance map recorded at the bias voltage of the +3 LL ($V_{sample}$ = 170 mV). **b,** Vertical line-cuts of the

conductance maps of the $0_+$, $0_-$, -1, +1, and +3 LLs along *A* and *B* atoms. The curves are offset vertically for clarity and the zero-line for these curves are denoted by dashed lines. **c,** The theoretical result showing amplitudes of these LLs on the *A* and *B* atoms. The amplitudes of the two components are calculated according to Eq. (2) and we use Guassian peaks to reflect the linewidth of the DOS at the *A* and *B* atoms.

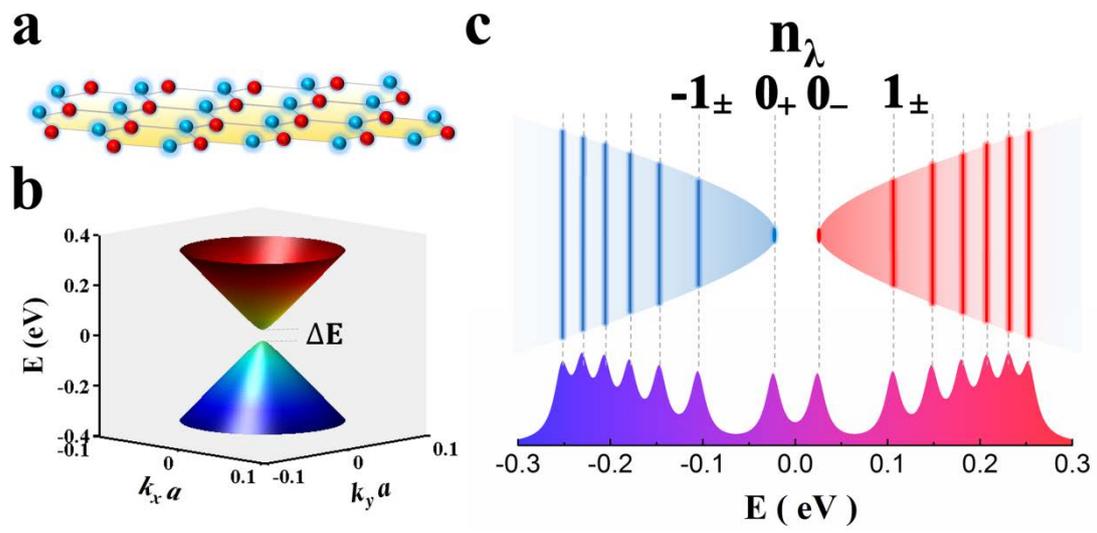

Figure 1

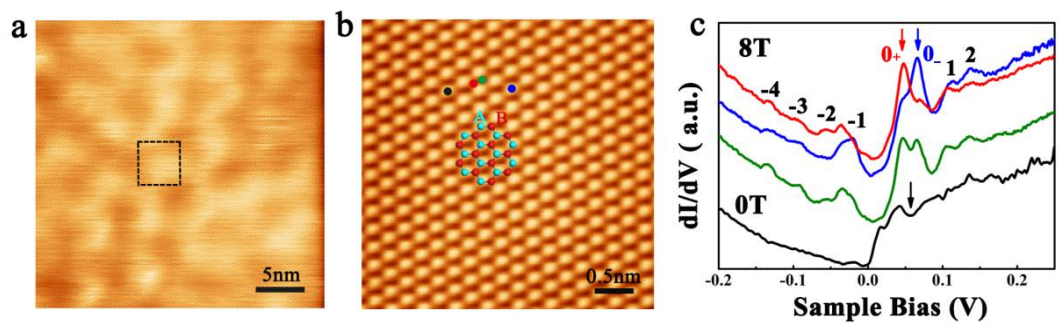

Figure 2

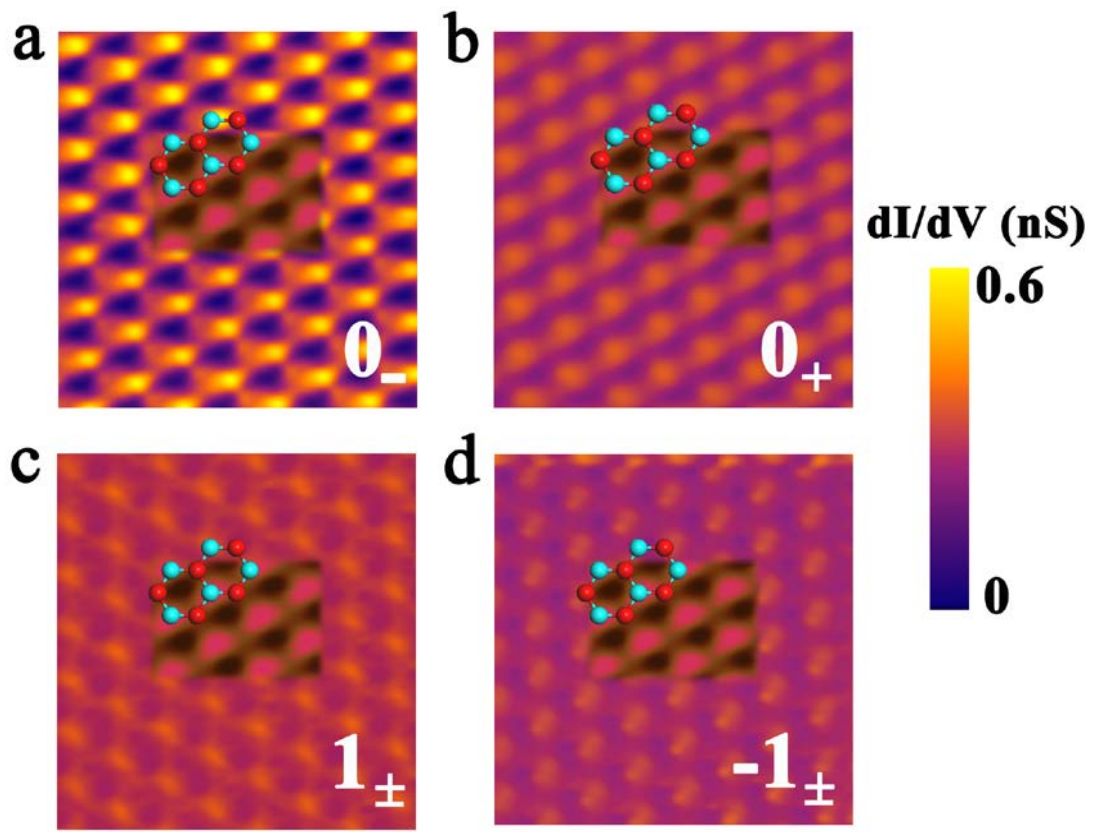

Figure 3

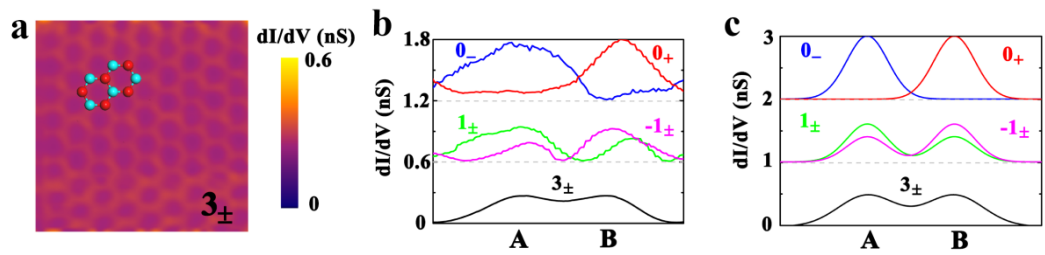

Figure 4